# Enhanced Formulae for Determining the Axial Behavior of

# Cylindrical Extension Springs

**Paredes, Manuel**

ICA, Université de Toulouse, UPS, INSA, ISAE-SUPAERO, MINES-ALBI, CNRS, 3 rue

Caroline Aigle, 31400 Toulouse, France

manuel.paredes@insa-toulouse.fr

**Stephan, Thomas**

CGR International

1 rue Frederic Joliot Curie 93274 SEVRAN

thomas.stephan@cgr-international.com

**Orcière, Hervé**

CGR International

Avenue jean moulin, Zone activité du plat, 05400 VEYNES

herve.orciere@cgr-international.com

**ABSTRACT**

Cylindrical extension springs have been commonly exploited in mechanical systems for years and their behavior could be considered as well identified. Nevertheless, it appears that the influence of the loops on the global stiffness is not yet taken into account properly. Moreover, it would be of key interest to analyze how initial tension in extension springs influences the beginning of the load-length curve. The paper investigates these topics using analytical, simulation and experimental approaches in order to help engineers design extension springs with greater accuracy. As a result, the stiffness of the loops has been analytically defined. It enables to calculate the global stiffness of extension springs with more accuracy and it is now possible to determine the effective beginning of the linear load-length relation of extension springs and thus to enlarge the operating range commonly defined by standards. Moreover, until now manufacturers had to define by a try and error process the axial pitch of the body of extension springs in order to obtain the required initial tension. Our study enables for the first time to calculate quickly this key parameter saving time on the manufacturing process of extension springs.

## 1. INTRODUCTION

Mechanical cylindrical springs are often used in mechanical devices for their ability to store and return energy. The range of applications is very wide. Cylindrical compression springs can be found in connectors [1], pump turbines [2] and extension springs can be found in robotic systems [1-3], static balancing machines [4-7] and medical applications [8-10] among others. Most of the industrial software available to help designers, such as Advanced Spring Design from SMI [11] and Spring Calculator Professional from IST [12] or websites dedicated to spring design [13-14] exploit common analytical formulae from the reference book on spring design written several years ago by Wahl [15]. Since that time, knowledge of extension springs has been improved by a few works on fatigue resistance [16-17], optimal design [18] and simulation using FEA [19]. Hopefully, recent works related to compression springs may also be of key interest for our study. Rodriguez [20], Paredes [21] and Dym [22] have performed experimental and analytical studies that may be successfully extended to cylindrical extension springs.

The present study focuses on the most common extension spring: the cold formed cylindrical extension spring with cylindrical wire. To increase readability, this will be referred to simply as an extension spring throughout this paper. The spring is composed of a cylindrical helix of n coils in contact (the body) : $L_k = d(n+1)$, plus a loop at each end ($L_0 = L_k + 2 L_H$), as illustrated in Fig. 1a.

Fig. 1 Load-length relation for extension springs, a: from European standards [23], b: from IST [24]

A designer expects an extension spring to have linear load-length behavior, which is mainly defined by the free length $L_0$, the initial tension $P_0$ and the spring rate k. This initial tension is obtained on the coiling machine by setting an axial pitch for the body that is lower than the wire diameter inducing prestress in the body of the spring. For a given length, $L_1$, of the system where the spring is inserted, the associated load, $P_1$, given by the spring can be calculated easily:

$$P_1 = k \, (L_0 - L_1) + P_0 \tag{1}$$

The accuracy of the calculated load depends directly on the accuracy of $L_0$, $P_0$ and k. Usually, the spring rate is calculated using the common formula:

$$k \; = \; \frac{G \, d^4}{8 \, n_a \, D^3} \tag{2}$$

where $n_a = n$ [13, 14, 23] or $n_a = n + 1$ to consider the stiffness of the loops [15, 24].

Thus, depending on the standard used, the stiffness of the loops can be neglected or approximated by adding one coil. As far as we are aware, no accurate formula exists to describe the stiffness of the loops and its effect on the global stiffness of the spring.

In addition, as shown by the dotted lines in Figure 1 (a) and b)), when the load is less than the initial tension, a small deflection exists. In fact, in this case, the initial tension maintains the coils of the body in contact but logically the loops, having no pretension, deflect. The spring exhibits a linear load-length relation only when all the initial tension inside the coils of the body is overcome. To be sure of having a linear load-length relation, the common rule for designers is to use an extension spring only between 20% and 80% of its theoretical range (i.e. from $L_0$ to $L_n$, where $L_n$ is defined by the Ultimate Tensile Strength of the material) [24]. This is a huge limitation and it would be of key interest to be able to estimate the beginning of the linear behavior in order to extend the range of applications of extension springs.

As far as we are aware, no study has been performed to analyze the load-length relation for extension springs with loads around the initial tension. Therefore, we propose a mixed study exploiting experimental, FEA and analytical aspects. This approach is illustrated throughout the paper on one extension spring with crossover loops. In section 2, we propose an experimental study of an extension spring to highlight its load-deflection relation at the beginning of deflection. Then, a simulation process using FEA is developed in section 3. The simulation takes the initial tension into account and highlights the relative contributions of the body and the loops to the global stiffness of the spring. Based on the remarks on observation of the simulation results, section 4 proposes an analytical approach, which is applied to the same spring, and conclusions are drawn in section 5.

## 2. EXPERIMENTAL OF EXTENSION SPRINGS

### 2.1 EXPERIMENTAL TEST BENCH

The experimental study was intended to highlight the deflection of the loops and its effect on the global stiffness. Thus a spring from a manufacturer's catalogue, with few active coils, was selected.

Details of the 1.4310 stainless steel spring [25] with crossover loops are given in table 1.

Table 1: details of the 1.4310 stainless steel spring

| d (mm) | $D_o$ (mm) | $L_o$ (mm) | n | $P_0$ (N) | k (N/mm) | $P_{max}$ (N) |
|--------|-----------|-----------|---|----------|----------|--------------|
| 1.5 | 7.1 | 17.9 | 5 | 26.2 | 50.4 | 151 |

Thus the common operating range (20%-80% [24]) for this spring is between 51.2 N and 126 N.

To estimate the accuracy of experiments and manufacturing tolerances, four specimens of this extension spring were tested on a Spring Test 1 test bench from Andilog [26] (see Fig. 2). The force gauge had a capacity of 500 N with 0.1% accuracy and 0.04 N resolution. The handle enabled a stroke of 2 mm per revolution. The displacement transducer with digital display had a resolution of 0.01 mm.

In order to test the springs in tension, a specific apparatus was employed with 3.4 mm diameter rods inside the loops (see Fig. 2). Because of this additional apparatus and because the displacement traducer is not directly close to the spring, it is necessary to correct the measure of  deflection S given by the transducer by taking into account the axial stiffness of the test bench. The corrected deflection $S_C$ for a given axial load F is thus: $S_C = S - F/K$. To evaluate the stiffness of the test bench K, the extension spring was replaced by a plain cylinder with 2 holes (see Fig.2).

Fig 2. Test bench

## 2.2 RESULTS

The results of the tests are presented in Fig. 3.

Fig 3. Experimental load-deflection curves for the four tested springs

Each spring clearly exhibits a bilinear load-deflection curve. It is thus possible to extract:

- the global rate k when the spring deflects fully (here from about 0.2 mm deflection)

- the initial tension $P_0$, which is calculated as the ordinate at x=0 of the line fitting the overall behavior

- the initial rate $k_i$ at the beginning of deflection when the coils are still in contact

The values obtained are detailed in Table 2.

Table 2: exploitation of the experimental results

|  | #1 | #2 | #3 | #4 | Mean | Standard deviation |
|---|---|---|---|---|---|---|
| global rate k (N/mm) | 46.4 | 47.2 | 48.5 | 47.2 | 47.3 | 0.9 |
| initial tension $P_0$ (N) | 24.8 | 26.1 | 28.1 | 27.1 | 26.5 | 1.4 |
| initial rate $k_i$ (N/mm) | 185.2 | 208.4 | 233.5 | 241.3 | 217.1 | 25.5 |

The four springs showed similar values of their global rate, with an average value of 47.3 N/mm, which is close to but lower than the value given in the catalogue (50.4 N/mm). The values of the initial tension were very close to that of the catalogue (26.2 N) and the average experimental value of $P_0$ was 26.5 N. We can also note that the transition load, $P_T$, between the initial behavior and the global behavior is greater than $P_0$ ($P_T = 33.9$ N for the average results). This confirms that the minimum operating load was greater than the initial tension. Nevertheless, the common rule for designers [24] induces a minimum operating load of 51.2 N. This value appears to be very conservative as the spring shows a linear load-deflection curve from $P_T = 33.9$ N. The linear operating range of this spring could reasonably be extended. We thus propose both FEA and analytical approaches to better estimate the transition load, $P_T$, of an extension spring.

## 3. SIMULATION OF THE AXIAL BEHAVIOR OF EXTENSION SPRINGS

### 3.1 SIMULATION PROCESS

3D Finite Element Analysis is commonly exploited to simulate the behavior of various springs [1, 27-32]. This kind of simulation could not be directly used in our study because, in the initial geometry of extension springs, the coils are in contact and that could cause some nodes to be merged. Moreover, the initial tension could not be taken into account easily. Alternatively, a simulation that integrates the shaping process of such a spring would make it possible to integrate the problem of contact and preload. Unfortunately, this would require a significant calculation time and would also require exploiting elastic-plastic deformation laws for the material that are not given neither by standards or wire manufacturers.

Known material properties are those of a 1.4310 stainless steel (EN 10270-3) and detailed in table 3.

Table 3: elastic parameters used on the finite element simulation

| Young Modulus (MPa) | Poisson ratio | Density (T/mm$^3$) |
| --- | --- | --- |
| 185 000 | 0.267 | 7.9e-9 |

To be able to simulate the initial tension, we adapted the simulation process suggested by Shimoseki [19], which uses three-dimensional, two-nodal beam elements. In this work, an extension spring is modeled by beam elements that fit the initial geometry of the spring and the initial tension is not considered. Later in the book, gap conditions are exploited to manage contact between coils in conical springs. We propose to combine the two approaches so as to be able to simulate the initial tension.

Hence, the extension spring is first modeled with a body having an axial pitch smaller than the wire diameter. The pitch value is determined depending on the required transition load $P_T$ (remember that $P_T$ is greater than the initial tension $P_0$) and the deflection of one coil using the following formula:

$$p \; = \; d - \frac{8 \, P_T \, D^3}{G \, d^4}$$

(3)

We can note that this pitch value is not only required to model the initial tension on the FE model but is also the pitch that is defined on the coiling machine so as to manufacture an extension spring with the appropriate initial tension. Until now, manufacturers had to define the axial pitch of the body of extension springs with a trial and error process. As far as we are aware, Eq. [3] is defined for the very first time and is thus of key interest for saving time in the manufacturing process of extension springs.

Then gaps are defined by using connectors in Abaqus with a stop function when the distance between nodes is equal to the wire diameter. One node of the upper loop is pinned and radial displacements of a node of the opposite loop are set equal to 0. Details of this model and of the simulation process are shown in Fig. 4. At the beginning of the simulation process (Fig. 4: initial), the spring exhibits a free length slightly lower than $L_0$ and there is no pretension. The spring is then extended at step 1 so that the distance between coils exceeds the wire diameter. Thus at the end of step 1, the length of the spring is greater than $L_0$. Then at step 2 the axial displacement previously set is deactivated so that the length decreases, the connectors are activated and the preload is set. At the end of step 2, the extension springs is preloaded (see Von Mises stress) and has the required free length $L_0$. On the next steps, an increasing axial load is applied in order to obtain a realistic load/deflection curve even for load values lower than the initial tension.

Fig 4. FEA of an extension spring with initial tension

The calculation time is less than 2 minutes on a using a laptop.

**3.2 FEA RESULTS**

The results of the simulation are presented in Fig. 5 and compared with the average experimental results for better readability.

Fig 5. Load-deflection curve of an extension spring with initial tension, FEA

The finite-element simulation shows very good correlation with the average experimental results. The global rate of the spring is 47.8 N/mm and the initial rate is 243 N/mm.

The FEA simulation also enables the behavior of the loops and the behavior of the body to be analyzed separately. To analyze the behavior of the coils, the stiffness of the body is artificially increased by taking the wire diameter to be 20 mm (instead of 1.4 mm) and, to analyze the behavior of the body, the stiffness of the loops is artificially increased by considering the wire diameter to be 20 mm. The results are presented in Fig. 5. We can see the expected full linear load-deflection curve for the loops but we can also see that the load-deflection curve of the body exhibits a bilinear curve with an initial rate inducing a deflection of approximately 0.5 coils. This initial deflection of the body has to be considered in an analytical approach to estimate the initial rate of extension springs.

## 4. ANALYTICAL STUDY

It is often of key interest for designers to be able to exploit analytical formulae instead of having to perform FEA. Analytical formulae are also useful for researchers to obtain kick evaluation processes for example in order to study dynamical issues in helical springs [33, 34]. To determine the global stiffness of an extension spring, the loops and the body can be studied separately.

### 4.1 STIFFNESS OF A CROSSOVER LOOP

In order to define the stiffness of a loop, it is first necessary to describe its geometry precisely. Fig. 6 details the geometry of a crossover loop.

Fig 6. Detailed geometry of a crossover loop

This geometry can be split into three parts:

1) a circular part linking the loop with the body and having a radius $R_{L1}$,

2) a circular part of radius $R_{L2}$ (commonly $R_{L2} = D/2$) at the end,

3) a linear part of length $L_L$ that links the two circular parts

For such geometry, Castigliano's second theorem [35] can be employed, as done by Dym for compression springs [22], to find the flexibility of a loop. It is obtained by adding the flexibilities of the three parts:

$$u_1\left(P\right) = P\, R_{L1} \int_0^{\frac{\pi}{2}} \left[ \frac{1}{GS} + \frac{\left(L_L\cos\theta_1 + R_{L1} - R_{L1}\sin\theta_1\right)^2}{GJ} + \frac{\left(L_L\sin\theta_1 + R_{L1}\cos\theta_1\right)^2}{EI} \right] d\theta_1$$

$$u_2\left(P\right) = P\, R_{L2} \int_0^{\pi} \left[ \frac{\cos^2\theta_2}{GS} + \frac{\sin^2\theta_2}{ES} + \frac{\left(R_{L2}\sin\theta_2\right)^2}{EI} \right] d\theta_2$$

$$u_3\left(P\right) = P \int_0^{L_L} \left[ \frac{1}{GS} + \frac{L_L^2 + x^2 - 2xL_L}{EI} \right] dx$$

This induces:

$$F_L = \frac{R_{L1}}{GJ}\left[ \frac{\pi}{4}L_L^2 + R_{L1}^2\left(\frac{3\pi}{4} - 2\right) + 2R_{L1}L_L \right] + \frac{\pi R_{L2}^3}{2EI} \tag{4}$$

with $J = \dfrac{\pi d^4}{64}$ ; $I = \dfrac{\pi d^4}{32}$ ; $L_L = \dfrac{D}{2} - R_{L1}$

## 4.2 GLOBAL STIFFNESS

It can be seen in Fig. 6 that the circular part with radius $R_{L1}$ decreases the number of coils of the body (that has the common helix shape). This decrease can be estimated by the following formula:

$$n_L \approx \frac{r_1}{\pi D} \qquad (5)$$

Thus the common formula for an extension spring (Eq. 2) can be adapted to define the flexibility of the remaining body:

$$F_b = \frac{8 \left( n - 2n_L \right) D^3}{G d^4} \qquad (6)$$

Eq. 4 and Eq. 6 can be combined to calculate the global rate of the spring, k:

$$k = \frac{1}{F_b + 2F_L} \qquad (7)$$

## 4.3 INITIAL STIFFNESS

At the beginning of deflection, the FEA analysis highlighted that the stiffness was due to the deflection of the loops and the deflection of half a coil of the body. Thus, Eq. 2 and Eq. 4 are combined to define the initial rate of the spring $k_i$:

$$k_i = \frac{1}{\dfrac{4 D^3}{G d^4} + 2F_L} \qquad (8)$$

## 4.4 COMPARISON WITH EXPERIMENTAL DATA

Several analytical approaches can be presented to estimate the behavior of an extension spring. For both models, $P_0$ equals 26.5 N and the material properties lead to G = 70000 MPa and E = 182000 MPa.

The experimental mean values are compared with the results of three analytical approaches: the common linear behavior with n active coils (k = 50.4 N/mm), the common linear behavior with n +1 active coils (k = 42.0 N/mm) and the proposed bi-linear behavior.

Table 4: experimental data

| | #1 | #2 | #3 | #4 | Mean | Standard deviation |
|---|---|---|---|---|---|---|
| global rate k (N/mm) | 46.4 | 47.2 | 48.5 | 47.2 | 47.3 | 0.9 |
| initial tension $P_0$ (N) | 24.8 | 26.1 | 28.1 | 27.1 | 26.5 | 1.4 |
| initial rate $k_i$ (N/mm) | 185.2 | 208.4 | 233.5 | 241.3 | 217.1 | 25.5 |

For the bilinear approach, the flexibility of the loops requires $R_{L1}$ to be defined. In the example presented here, $R_{L1}$ = 1.5 mm. This induces, k = 46.2 N/mm; $k_i$ = 231 N/mm. The three analytical approaches are compared with the experimental results in Fig 7.

Fig 7. Load-deflection curve of an extension spring with initial tension, analytical

We can see that the proposed bilinear approach gives results that are very close to the experimental ones. Thanks to the evaluation of the behavior of the loops, the global stiffness of the bilinear model is more accurate than that found with the linear approaches commonly used, which tend to underestimate or overestimate the global stiffness.

Moreover, the common operating range (20%-80% [24]) for this spring is between 51.2 N and 126 N. The proposed bilinear approach enables the transition to be estimated using Eq. 9.

$$P_T = P_0 \frac{k_i}{k_i - k} = 33.1 N \tag{9}$$

Thus the lower bound of the operating range can be reduced (from 51.2 N to 33.1 N), significantly increasing the potential operating range of the spring [6%-80%].

In addition, with this calculated value of $P_T$ it is possible for manufacturer to determine the needed pitch on the coiling machine and to reduce set up time.

## 5. CONCLUSIONS

The global stiffness of extension springs is commonly calculated using the formula dedicated to compression springs and two options are proposed for defining the active coils $n_a$. The first one defines $n_a$ as the number of coils of the body: $n_a = n$, and the second one considers the influence of the loops by adding one coil: $n_a = n + 1$. The aim of the study was, first, to evaluate the accuracy of these formulae and propose new approaches by performing experimental, finite-element and analytical analyses.

The experimental study clearly showed what is only suggested by standards: the initial tension induces the extension spring to exhibit a bilinear load-deflection curve with high stiffness at the beginning of deflection.

To analyze the behavior of extension springs in greater depth, a finite-element study was then developed. As far as we are aware, this modeled the initial load of extension springs for the first time. This was done using beams for the spring and connectors to introduce the pretension. Simulations highlighted that the initial rate of the spring was due to the deflection of the loops and about half a coil of the body. Thanks to these results, an analytical approach was developed using Castigliano's second theorem and a bilinear behavior was thus modeled. This new model gives a more accurate evaluation of the global stiffness of the extension spring. Moreover, it enables the transition load to be evaluated (from the initial stiffness to the global stiffness). This transition load can opportunely be exploited to enlarge the common operating range of extension springs, which currently appears to be very conservative.

Finally, this study can also be directly exploited by manufacturers to quickly calculate the pitch to be set on the coiling machine in order to obtain the expected initial tension on the load-deflection curve, saving time on the manufacturing process.

Of course, it is important to bear in mind that this study was illustrated using a single spring geometry. It can thus be considered as a first, but important, step in assessing the quality of the analytical formulae concerning extension springs. Several studies (using experimental/FEA/analytical approaches) were performed on springs having other geometries and indicated the same conclusions but further investigations will be necessary to fully cover the design space. It could also be interesting to analyze the stress in the body and the stresses in the coils using 3D FEA in order to give a more accurate estimate of the upper bound of the operating range of extension springs.

**NOMENCLATURE**

| | |
|---|---|
| $c$ | spring index = D/d |
| $d$ | wire diameter |
| $D$ | mean diameter of the spring |
| $D_o$ | external diameter of the spring |
| $E$ | Young's modulus |
| $G$ | torsional modulus |
| $k$ | global spring rate |
| $K_i$ | initial spring rate |
| $L_0$ | free external length |
| $L_L$ | length of the linear part of the loop |
| $L_1$ | length related to $P_1$ |
| $L_2$ | length related to $P_2$ |
| $n$ | number of coils of the body |
| $n_a$ | number of active coils |
| $P$ | axial load |
| $P_0$ | initial tension load related to $L_0$ |
| $P_1$ | axial load related to $L_1$ |

$P_2$        axial load related to $L_2$

$P_T$        axial transition load between the initial behavior and the global behavior

$R_{L1}$       radius

$R_{L2}$       radius

$\theta_1$        angle

$\theta_2$        angle